\documentclass[aps,prb,twocolumn,superscriptaddress,nofootinbib,a4]{revtex4}
\usepackage{graphicx}
\usepackage{latexsym}
\usepackage{array}
\usepackage{color}
\newcolumntype{P}[1]{>{\centering\arraybackslash}p{#1}}
\newcolumntype{M}[1]{>{\centering\arraybackslash}m{#1}}
\input{epsf}
\begin{document}
\title{Josephson coupling in the dissipative state of a thermally hysteretic $\mu$-SQUID}
\author{Sourav Biswas}
\affiliation{Department of Physics, Indian Institute of Technology Kanpur, Kanpur 208016, India}
\author{Clemens B. Winkelmann}
\affiliation{\mbox{Univ.} Grenoble Alpes, CNRS, Grenoble INP$^\dagger$, Institut N\'eel, 38000 Grenoble, France}
\author{Herv\'e Courtois}
\affiliation{\mbox{Univ.} Grenoble Alpes, CNRS, Grenoble INP$^\dagger$, Institut N\'eel, 38000 Grenoble, France}
\author{Anjan K. Gupta}
\email[]{anjankg@iitk.ac.in}
\affiliation{Department of Physics, Indian Institute of Technology Kanpur, Kanpur 208016, India}
\date{\today}

\begin{abstract}
Micron-sized superconducting interference devices ($\mu$-SQUIDs) based on constrictions optimized for minimizing thermal runaway are shown to exhibit voltage oscillations with applied magnetic flux despite their hysteretic behavior. We explain this remarkable feature by a significant supercurrent contribution surviving deep into the resistive state, due to efficient heat evacuation. A resistively shunted junction model, complemented by a thermal balance determining the amplitude of the critical current, describes well all experimental observations, including the flux modulation of the (dynamic) retrapping current and voltage by introducing a single dimensionless parameter. Thus hysteretic $\mu$-SQUIDs can be operated in the voltage read-out mode with a faster response. The quantitative modeling of this regime incorporating both heating and phase dynamics paves the way for further optimization of $\mu$-SQUIDs for nano-magnetism.
\end{abstract}

\maketitle
\section{Introduction}
A superconducting quantum interference device (SQUID), in which two Josephson junctions form a closed loop, exhibits a modulation of the critical current $I_{\rm c}$, as a function of the magnetic flux through the loop with a period $\Phi_{\rm 0}=h/2e$. It is the most sensitive magnetic field transducer to date.\cite{granatasquid,recentsquidreview} Miniaturized SQUIDs have been used for probing magnetic properties of nanoparticles \cite{mag, mag1, cobultnano} and surfaces with sub-$\mu$m resolution.\cite{scanningsquidmicro1, scanningsquidmicro2} The coupling of a nano-particle's magnetic flux to a $\mu$- or nano-SQUID is far better \cite{granatasquid} than to a conventional SQUID, leading to a magnetic moment resolution down to below 1 $\mu_B$.\cite{mag1,mag2} Thus, optimizing such $\mu$-SQUIDs in terms of sensitivity, ease of fabrication and operation, or operating temperature and magnetic field range is the focus of a large panel of recent works.\cite{highqualitysquid, siliconsquid, nbnanosquid, diamondsquid, lownoisesquid, NbNsquid, giazotto} However, hysteresis in the current-voltage characteristics (IVCs) of $\mu$-SQUIDs severely limits their flux resolution and speed. Hysteresis in conventional SQUIDs based on superconductor-insulator-superconductor Josephson junctions is well understood in the frame of the resistively and capacitively shunted junction (RCSJ) model.\cite{squidbook,tinkhambook} In contrast, hysteresis in weak-link (WL) based $\mu$-SQUIDs arises from the Joule heating leading to a self-sustained hot-spot.\cite{scocpol, tinkhamnanowire, herveprl, hazraprb} Various strategies to avoid this phenomenon have been explored\cite{baranovprb, nikhilsust, lahmapl} but with incomplete success. Recently, high sensitivity non-hysteretic nano-SQUIDs, based on high critical temperature cuprate grain boundary junctions shunted with gold films, have been demonstrated for nano-magnetism.\cite{hightc1,hightc2} Although such devices have some advantage as heating effects seem to be negligible there, their fabrication is quite challenging.

In a WL biased with a current $I$ close to the critical current $I_{\rm c}$, the transition to the dissipative state is triggered by a phase slip\cite{gregory,nayanaprl}, which changes the quantum phase difference $\varphi$ between the leads by $2\pi$. The ensuing voltage peak, and thus heating, generally suffices to create an avalanche of phase-slips, driving the local WL temperature $T_{\rm WL}$ above the bath temperature $T_{\rm b}$. When the bias current is ramped down, superconductivity is recovered only at the so-called retrapping current ($<I_{\rm c}$) leading to hysteresis. The hot-spot model by Skocpol, Beasley and Tinkham (SBT)\cite{scocpol} considers that in the finite-voltage state the temperature $T_{\rm WL}$ is above the critical temperature $T_{\rm c}$. The Josephson coupling is then lost so that no SQUID-type behavior is observed in this state for most of the temperature range.\cite{nikhilprl,hazrair,hazraprb} Still, it has been observed in some devices based on WLs\cite{wangjap,tinkhamnanowire,wangapl} and SNS junctions.\cite{anger,ronzani,krasnov} In most cases, the theoretical modeling neglected thermal effects and relied on a conventional RCSJ model, but with an effective capacitance\cite{song} well above the actual geometric one. Eventually, the SBT model was also extended to the case of a WL temperature remaining below $T_{\rm c}$, still ignoring the WL phase dynamics.\cite{wangjap,wangapl}

Recently, some of us proposed a dynamic thermal model\cite{anjanjap} of WLs, incorporating both the overheating of the WL at a temperature $T_{\rm WL}$ and a resistively-shunted junction (RSJ) type phase dynamics. If $T_{\rm WL}$ remains below $T_{\rm c}$ then the Josephson coupling across the WL is not fully destroyed. The Josephson current, together with the normal current, persists over a portion of the finite voltage branch of the IVCs,\cite{twofluidmodel} thus leading to SQUID-type voltage oscillations. This dynamic thermal model (DTM) is in between the SBT model, where the Josephson coupling does not exist at non-zero voltages, and the R(C)SJ model that ignores the thermal effects. A similar approach was used to describe the radio-frequency response of a SNS junction.\cite{PRB-DeCecco} Alternatively, hysteresis in the phase slip-controlled regime can be described using a more elaborate non-equilibrium approach using time-dependent Ginzburg-Landau equations.\cite{peeter}

In this paper, we report temperature- and magnetic field-dependent transport in $\mu$-SQUIDs with a geometry optimized for a moderate Joule heating. Despite being clearly hysteretic, the devices exhibit voltage oscillations with the magnetic flux. The related non-zero flux-sensitive supercurrent contribution surviving well above the critical current, which cannot be understood using the static SBT model, is analyzed with the dynamic thermal model, which quantitatively captures every observation. We eventually discuss the flux-sensitivity of the studied hysteretic $\mu$-SQUIDs in the dynamic regime.

\section{Dynamic Thermal Model}

In the dissipative state, phase slips\cite{nayanaprl,bezryadin-phase-slip,Bezryadin} occur at a rate $V/\Phi_{\rm 0} \simeq \tau_{\rm J}^{\rm -1} = R_{\rm N}I_{\rm c}^{\rm 0}/\Phi_0$, where $I_{\rm c}^{\rm 0}$ is the zero-magnetic field critical current, taken here at the bath temperature. Each phase-slip deposits a Joule heat $I\Phi_{\rm 0}$ leading to a temperature rise in the WL region. The Joule heat is generated over a length scale determined by the inelastic quasi-particle diffusion length,\cite{SBT-phase-slip,dolan-PS} much longer than the WL dimensions studied here. Thus we assume a uniform\cite{nayanaprl} temperature over the entire WL, which determines its critical current $I_{\rm c}$. The characteristic time for the thermal balance is $\tau_{\rm th} = C_{\rm WL}/k \sim$ where $C_{\rm WL}$ is the WL heat capacity and $k$ is the thermal conductance to the bath. Under the quasi-static approximation, the instantaneous WL temperature $\mathcal{T_{\rm WL}}$ is dictated by a thermal balance between the Joule heat and the conduction to the bath $k(\mathcal{T_{\rm WL}}-T_{\rm b})$.

If the temperature $\mathcal{T_{\rm WL}}$ remains below $T_{\rm c}$ at any instant of time $t$, the bias current $I$ is dynamically shared between a super-current $\mathcal{I}_{\rm s}(t)$ and a complementary normal current. This gives rise to a time-dependent voltage $\mathcal{V}(t)=R_{\rm N}(I-\mathcal{I}_{\rm s}(t))$ related to the phase difference $\varphi$ through the Josephson relation $\mathcal{V}(t)=\Phi_{\rm 0} \dot{\varphi} /2\pi$. The heat balance equation governing the dynamics of temperature is written as $C_{\rm WL}\dot{\mathcal{T_{\rm WL}}}+k(\mathcal{T_{\rm WL}}-T_{\rm b})=\mathcal{V}^{2}(t)/R_{\rm N}$. These two equations can be re-arranged in terms of dimensionless variables as:\cite{anjanjap}
\begin{equation}
\dot{\varphi}=2\pi(i-i_{\rm s}),
\label{eq:current}
\end{equation}
\begin{equation}
\alpha \dot{p}+p=\frac{\beta}{4\pi^2}\dot{\varphi}^{2}
\label{eq:temp}
\end{equation}
Here, the currents $i$ and $i_{\rm s}$ are, respectively, the currents $I$ and $\mathcal{I}_{\rm s}$ in units of the zero-magnetic field critical current $I_{\rm c}^{\rm 0}(T_{\rm b})$. The time unit is $\tau_{\rm J}$ and $\alpha=\tau_{\rm th}/\tau_{\rm J}$. The dimensionless temperature is defined as:
\begin{equation}
p =\frac{\mathcal{T_{\rm WL}}- T_{\rm b}}{T_{\rm c}-T_{\rm b}}.
\end{equation}
We also define the dimensionless parameter
\begin{equation}
\beta =\frac{R_{\rm N}{I_{\rm c}^{\rm 0}}^2(T_{\rm b})}{k(T_{\rm c}-T_{\rm b})}
\label{eq:beta}
\end{equation}
as the ratio of Joule heat generation (at $I_{\rm c}^{\rm 0}$) and heat evacuation (at $T_{\rm c}$).

\begin{figure}[t!]
	\includegraphics[width=\columnwidth]{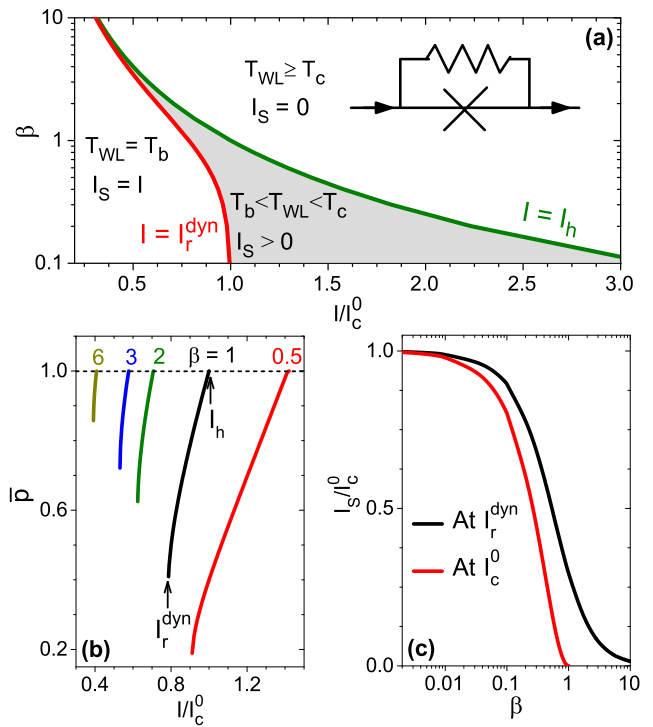}
	\caption{Device state diagram with the grey shaded region as the dynamic regime where the WL has a finite voltage but carries a non-zero supercurrent. The region on the left of this grey-region is the zero-voltage state where all the bias current is carried as supercurrent while the right-region has no supercurrent. The red (green) line depicts the $\beta$ dependence of $I_{\rm r}^{\rm dyn}$ ($I_{\rm h}$). Inset: Equivalent circuit of the DTM. (b) Variation of the dimensionless time-averaged WL temperature $\overline{p}$ with the bias current in the dynamic regime for different $\beta$ values. (c) Ratio of the time-averaged supercurrent $I_{\rm s}$ and the critical current $I_{\rm c}^0$ as a function of $\beta$ at the bias current values of $I_{\rm r}^{\rm dyn}$ and $I_{\rm c}^{\rm 0}$.}
	%}
	\label{fig:model}
\end{figure}

The static retrapping current \cite{notation} $I_{\rm h}$ is defined by the WL being right at the critical temperature ($\mathcal{T_{\rm WL}}=T_{\rm c}$). From \mbox{Eq. \ref{eq:temp}}, the thermal balance $k(\mathcal{T_{\rm WL}}-T_{\rm b}) = R_{\rm N}I^{\rm 2}$ gives:
\begin{equation}
I_{\rm h}=I_{\rm c}^{\rm 0}/\sqrt{\beta}.
\end{equation}
At a larger current $I> I_{\rm h}$ we have $\mathcal{T_{\rm WL}}>T_{\rm c}$ so that there is no Josephson coupling and hence $V=R_{\rm N}I$. A non-zero supercurrent can be carried by the WL only for $I<I_{\rm h}$.

The dynamic retrapping current $I_{\rm r}^{\rm dyn}$ is the current below which the dynamic state ceases to exist and the zero-voltage state becomes stable against any phase-slip. Here and in the following, we consider WLs featuring a linear temperature dependence of the critical current and a sinusoidal current-phase relation $\mathcal{I}_{\rm s}(\varphi)$. One can then obtain by solving \mbox{Eqs. (\ref{eq:current},\ref{eq:temp})}:\cite{anjanjap}
\begin{equation}
2\beta^2 \left[\frac{I_{\rm r}^{\rm dyn}}{I_{\rm c}^{\rm 0}}\right]^2=\sqrt{1+4\beta^2}-1.
\label{dynamic}
\end{equation}
From numerical simulations, we find that a non-sinusoidal $\mathcal{I}_{\rm s}(\varphi)$, within a regime of single-valued current, negligibly affects this relation. Moreover, the elevated WL temperature in the dynamic state gives rise to an increase in coherence length $\xi$ and thus the current-phase relation $\mathcal{I}_{\rm s}(\varphi)$ is close to sinusoidal.

The extent of the dynamic regime, defined by the current bias window $I_{\rm r}^{\rm dyn}< I < I_{\rm h}$, depends on the value of the dimensionless parameter $\beta$. Figure \ref{fig:model}(a) depicts the device state diagram found using the $\beta$ dependence of $I_{\rm r}^{\rm dyn}$ and $I_{\rm h}$. For large values of $\beta$, \mbox{i.e.} poor heat evacuation and/or high $I_{\rm c}^{\rm 0}$, $I_{\rm r}^{\rm dyn}$ and $I_{\rm h}$ are both below $I_{\rm c}^{\rm 0}$ and very close to each other. The dynamic regime then occurs in a bias current window of vanishing width, making its observation in IVCs practically impossible.\cite{nikhilprl,nikhilsust} In this limit, the physics is well described by the SBT and other static thermal models.\cite{scocpol,tinkhamnanowire,nikhilprl} In contrast, for $\beta$ of about unity or smaller, the dynamic regime spans over a significant current range. The static retrapping current $I_{\rm h}$ can then significantly exceed $I_{\rm c}^{\rm 0}$ and the dynamic retrapping current $I_{\rm r}^{\rm dyn}$ is close to $I_{\rm c}^{\rm 0}$. For extremely good heat evacuation $\beta \to 0$, one has $I_{\rm h}\to \infty$, $I_{\rm r}^{\rm dyn}\to I_{\rm c}^{\rm 0}$ and RSJ model is recovered.

In the dynamic regime between $I_{\rm r}^{\rm dyn}$ and $I_{\rm h}$, the WL temperature $\mathcal{T_{\rm WL}}$ oscillates with time about an average value. However, for a large value of $\alpha$, the magnitude of these oscillations is negligible compared to the average WL temperature.\cite{anjanjap} This is always the case as $\tau_{\rm th}$, which can range from tens of ns to $\mu$s, is greater than $\tau_{\rm J}$, which is of ps order. The WL can thus be considered at a constant (time-averaged) temperature $T_{\rm WL}$.\cite{anjanjap,PRB-DeCecco,nayanaprl} The corresponding time-averaged reduced temperature $\overline{p}$ can be calculated as a function of the current bias $i$ from \mbox{Eqs. (\ref{eq:current}, \ref{eq:temp})}:\cite{anjanjap}
\begin{equation}
\overline{p}=\frac{i^2\beta^2+\sqrt{-i^2\beta^2+i^4\beta^2+i^6\beta^4}}{1+i^2\beta^2},
\label{temperature_p}
\end{equation}
with $\beta$ as the single parameter. \mbox{Figure \ref{fig:model}(b)} shows how $\overline{p}$ (or equivalently $T_{\rm WL}$) decreases with the current bias for various values of the parameter $\beta$, starting from 1 (or $T_{\rm c}$) at $I_{\rm h}$. At every bias, one can thus calculate the critical current $I_{\rm c}(T_{\rm WL})$, and the related time-averaged voltage as:
\begin{equation}
V=R_{\rm N}\sqrt{I^2-I_{\rm c}^2(T_{\rm WL})},
\end{equation}
and the time-averaged supercurrent $I_{\rm s}$ as:
\begin{equation}
I_{\rm s}=I-V/R_{\rm N}=I-\sqrt{I^2-I_{\rm c}^2(T_{\rm WL})}.
\end{equation}
Figure \ref{fig:model}(c) shows $I_{\rm s}$ as a function of the parameter $\beta$ at bias current values equal to $I_{\rm r}^{\rm dyn}$ and $I_{\rm c}^{\rm 0}$. For $\beta > 1$, the time-averaged supercurrent $I_{\rm s}$ is zero when the bias current reaches $I_{\rm c}^{\rm 0}$.\cite{scocpol,nikhilprl} In practice, as soon as $\beta$ exceeds about 2, the WL switches almost immediately to a fully normal state with $T_{\rm WL}\geq T_{\rm c}$ and (almost) zero supercurrent. For small $\beta$ values, the supercurrent $I_{\rm s}$ is comparable to the full critical current $I_{\rm c}^{\rm 0}$.

\section{Experimental Details}
The fabrication of the $\mu$-SQUIDs starts with the deposition of a Nb thin film with thickness 40 nm on a Si substrate. A resist layer was afterwards patterned using laser lithography for the outer leads and contact pads, and electron-beam lithography for the smaller structures. A 25-nm thick Al layer was then deposited followed by lift-off. The latter acting as a protective mask, the Nb devices were obtained by a SF$_{\rm 6}$ reactive-ion etch. The Al mask is eventually chemically etched. Figure \ref{fig:ivc}(b) shows the loop of a SQUID with the two constrictions, of nominal width and length 40 and 160 nm respectively, in parallel. The critical current $I_{\rm c}$ was tuned down (to the 100 $\mu$A range) by trimming\cite{lahmapl} down the Nb thickness to 20 ($\pm$ 2) nm in subsequent reactive-ion etching steps without Al mask. The etching process can also lead to a reduction in $T_{\rm c}$ due to the appearance of a non-superconducting layer on top, bottom and also on the sides.\cite{gubin,siegel,physicac} However the thickness of such a layer\cite{cooper,proximity} is estimated to be $\sim$ 2 nm only, which leaves a large and effective superconducting channel at the core of the WLs.

Electrical transport measurements were carried out in a closed-cycle refrigerator with base temperature 1.3 K. The electrical signals are thoroughly filtered, both at room temperature ($\pi$ filters) and at base temperature (copper-powder filters). Home-made ground-isolated current sources and voltage amplifiers were used. From the temperature-dependent four-probe transport measurements of a first device, we find the onset of superconductivity at 8.6 K and a sheet resistance $R_\Box$ = 5.8 $\Omega$ in the normal state. In the following, we will present experimental data mainly from one sample. Another sample featured a similar behavior, see Appendix A.

\section{Results}
\begin{figure}[t!]
	\includegraphics[width=\columnwidth]{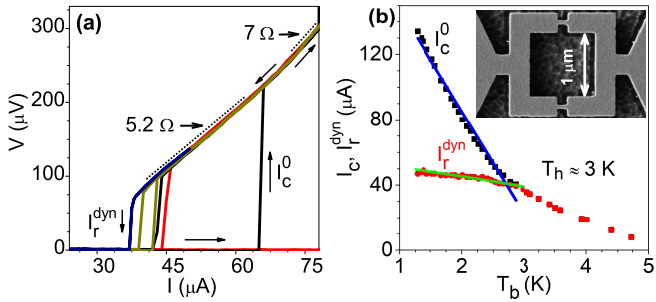}
	\caption{(a) Current-voltage characteristics at different temperatures in the hysteretic regime. Red, black, yellow and blue colors correspond to $T_{\rm b}$ = 1.67, 2.27, 2.77 and 2.98 K, respectively. (b) Temperature dependence of the critical current $I_{\rm c}^{\rm 0}$ and the (dynamic) retrapping current $I_{\rm r}^{\rm dyn}$. The crossover temperature $T_{\rm h}$ is close to 3 K. The solid lines are fits below $T_{\rm h}$. The critical current $I_{\rm c}^{\rm 0}$ is fit to 211.2$(1-T_{\rm b}/3.3)$, which gives the zero-temperature critical current $I_{\rm c0}=$ 211.2 $\mu$A. The dynamic retrapping $I_{\rm r}^{\rm dyn}$ is fit to Eq. \ref{dynamic}, which gives the WL critical temperature $T_{\rm c}$ = 6.0 K. Inset: Scanning electron micrograph of the SQUID loop.}
	\label{fig:ivc}
\end{figure}

Zero-field current-voltage characteristics (IVCs) of a $\mu$-SQUID at various bath temperatures $T_{\rm b}$ are shown in \mbox{Fig. \ref{fig:ivc}(a)}. The critical current $I_{\rm c}^{\rm 0}$ varies strongly with temperature and exceeds 100 $\mu$A below 2 K, see \mbox{Fig. \ref{fig:ivc}(b)}. The critical current density ($J_{\rm c}$) at 1.3 K, found as 21.1 MA/cm$^2$, is close to the Ginzburg Landau depairing current density $J_{\rm dp} = (2/3)^{3/2}[H_{\rm c}(0)/\lambda]$ = 36 MA/cm$^2$ (at zero temperature), estimated using parameters for bulk and clean Nb.\cite{tinkhambook} This $J_{\rm c}$ value is similar to our earlier devices.\cite{nikhilprl} Together with the linear dependence (up to $T_{\rm h}$) of the critical current $I_{\rm c}^{\rm 0}$ with the bath temperature $T_{\rm b}$, this confirms the intrinsic superconducting nature of the WLs as opposed to that of SNS WLs.\cite{SNS_Ic,PRB-DeCecco,nikhilprl} Below the crossover temperature $T_{\rm h}\approx3$ K, hysteresis is observed with a retrapping to the zero-resistance state at a well-defined \textit{dynamic} re-trapping current \cite{anjanjap} $I_{\rm r}^{\rm dyn}$, see \mbox{Fig. \ref{fig:ivc}(a)}. A fit of its temperature dependence to Eq. \ref{dynamic} provides the value for the critical temperature of the WL $T_{\rm c}$ = 6.0 K. At bias currents significantly above $I_{\rm c}^{\rm 0}$, further thermal instabilities occur in larger portions of the device as evidenced from additional re-trapping currents, see Appendix A. Our study here is focussed on the hysteretic regime and at bias currents below or in the vicinity of the critical current $I_{\rm c}$.

In the dissipative branch, the IVC slope $dV/dI$ varies with the bias current from 5.2 $\Omega$ (just above $I_{\rm r}^{\rm dyn}$) to 7 $\Omega$ (above $I_{\rm c}^{\rm 0}$) at 1.67 K, see \mbox{Fig. \ref{fig:ivc}(a)}. In the low-bias regime of interest here, the differential resistance is always close to $R_{\rm N}=5.2$ $\Omega$ independent of bias current and temperature, see Appendix A. This value is significantly below the value of 11.5 $\Omega$ for the resistance of two WLs in parallel estimated from the sheet resistance. This indicates that the WLs are not fully resistive in this dissipative state.  

\begin{figure}[t!]
	\includegraphics[width=\columnwidth]{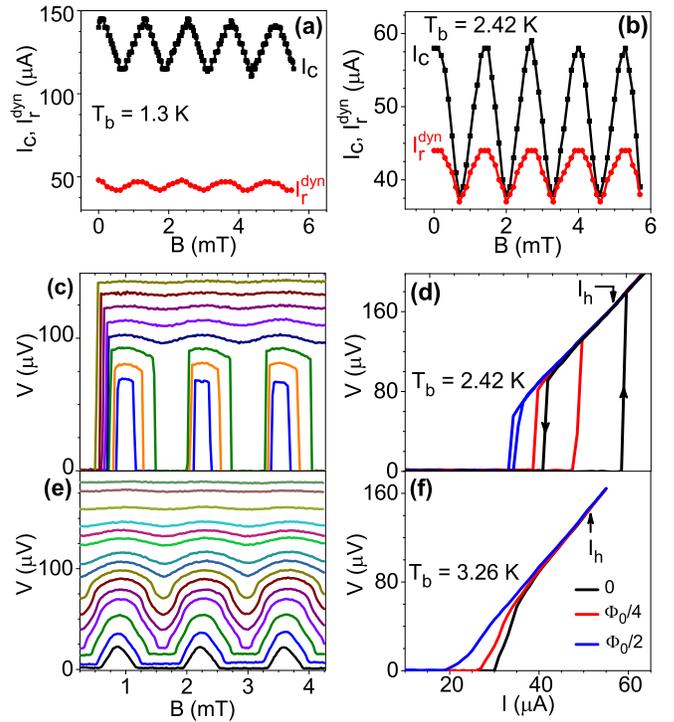}
	\caption{(a,b) Oscillations of the critical and the dynamic retrapping currents with the magnetic field in the hysteretic regime at 1.3 and 2.42 K, respectively. (c) Voltage oscillation with magnetic field for $I$ = 39 to 53 $\mu$A with 2 $\mu$A intervals at 2.42 K. (e) The same for $I$ = 25 to 50 $\mu$A at 3.26 K. (d,f) IVCs at different flux values ($0, \Phi_{\rm 0}/2$ and $\Phi_{\rm 0}/4$) at 2.42 and 3.26 K, respectively.}
	\label{fig:bvosc}
\end{figure}

When applying a perpendicular magnetic field $B$, and thus a flux $\Phi=B.S$, the critical current $I_{\rm c}$ displays a $\Phi_{\rm 0}$-periodic modulation, taking an effective SQUID loop area S = 1.6 $\mu$m$^2$. The flux modulation of the critical current, starting from its maximal zero-field value $I_{\rm c}^{\rm 0}$, is not complete and has a rather triangular shape, see \mbox{Figs. \ref{fig:bvosc}(a,b)}. This behavior cannot be explained solely by asymmetric critical currents between the two arms. Moreover, self-flux related effects,\cite{squidbook} related to the loop inductance $L$, are negligible here, as we estimate $LI_{\rm c}^{\rm 0}/\Phi_{\rm 0} <$ 0.1. For a WL with a length $\ell\geq\xi$, the supercurrent-phase $\mathcal{I}_{\rm s}(\varphi)$ relation is non-sinusoidal.\cite{wolf,hazrananosquid,iphi1,likharev} Numerical calculations using a non-sinusoidal $I_{\rm s}(\varphi)$ relation indeed yield an incomplete cancellation of $I_{\rm c}$ at $\Phi=\Phi_{\rm 0}/2$ (Appendix B), similar to the experimental behavior.

Strikingly, the retrapping current $I_{\rm r}^{\rm dyn}$ also shows oscillations with the magnetic flux, see \mbox{Figs. \ref{fig:bvosc}(a,b)}, in contradiction with the SBT picture of a fully normal state of the device in the dissipative state. A similar feature was observed in SQUIDs based on normal metal WLs \cite{anger,ronzani,krasnov} but not satisfactorily explained. It constitutes a first indication that superconductive coupling is not fully suppressed by the electron heating in the dissipative branch of our hysteretic devices.

\begin{figure}[t!]
	\includegraphics[width=\columnwidth]{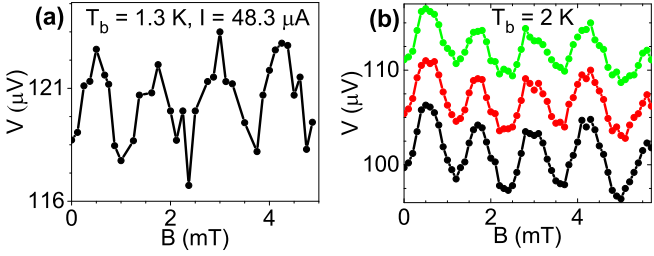}
	\caption{(a) $V-B$ oscillation at the lowest temperature of 1.3 K at a bias current 48.3 $\mu$A just above the dynamic retrapping current $I_{\rm r}^{\rm dyn}$. (b) Same at 2 K for three different bias currents (45, 46 and 47 $\mu$A) very close to $I_{\rm r}^{\rm dyn}$. At these lower temperatures, voltage values are extracted from IVCs at different magnetic field values.}
	\label{fig:bv1p3k}
\end{figure}

Moreover, the SQUID voltage also shows an oscillatory dependence on magnetic field for a wide bias current window, see \mbox{Figs. \ref{fig:bvosc}(c-f)}. This is completely opposed to the SBT model behavior, in which the dissipative state displays no signature of Josephson coupling across the SQUID WLs. The initial jump in voltage, seen for $I_{\rm r}^{\rm dyn}<I<I_{\rm c}$ (see \mbox{Fig. \ref{fig:bvosc}(c))}, occurs due to the first arrival to the resistive branch. At a fixed temperature, the IVCs at different flux-values are found to merge on the linear branch, see \mbox{Fig. \ref{fig:bvosc}(d,f)}, beyond a particular bias current. We identify this current as the {\em static} retrapping current $I_{\rm h}$,\cite{anjanjap} as discussed in the model section. The $V-B$ oscillations consistently disappear at a bias current beyond $I_{\rm h}$. At lower temperatures ($T_{\rm b}$ = 1.3 and 2 K) in the hysteretic regime, $V-B$ oscillations are observed over a narrow bias current span just above $I_{\rm r}^{\rm dyn}$, see  \mbox{Fig. \ref{fig:bv1p3k} (a,b)}.

The flux-to-voltage transduction function $V_\Phi$, defined as the maximum of $\partial V/\partial \Phi(\Phi)$, is found to be 27 $\mu$V/$\Phi_{\rm 0}$ in the dynamic regime at $T_{\rm b}$ = 2.42 K. With a voltage noise of $\sim$ 1 nV$/\sqrt{\rm Hz}$ in our circuit, this gives a flux noise density $\sqrt{S_\Phi} \approx$ 37 $\mu \Phi_{\rm 0}/\sqrt{\rm Hz}$. In the non-hysteretic regime, thanks to higher flux-to-voltage transduction $V_\Phi$, the sensitivity $\sqrt{S_\Phi}$ reaches 3 $\mu \Phi_{\rm 0}/\sqrt{\rm Hz}$ at $T_{\rm b}$ = 3.26 K. The latter value is similar to the ones reported in non-hysteretic $\mu$-SQUIDs with room temperature amplifiers.\cite{granatasquid,troeman-2007}

\section{Discussion}
For modeling the $\mu$-SQUID, we assume both WLs to be identical, with a temperature-dependent critical current $I_{\rm c}^{\rm 0}/2$, a temperature-independent normal-state resistance $2R_{\rm N}$, and a heat-loss coefficient $k/2$. The two WLs' phases $\varphi_{\rm 1}$ and $\varphi_{\rm 2}$ maintain a constant difference $\varphi_{\rm 1}- \varphi_{\rm 2}=2\pi \Phi/\Phi_{\rm 0}$, forced by the magnetic flux through the SQUID loop. Consequently, the two WLs' average temperatures are identical in the dynamic steady state. Eventually, the SQUID behaves as a single WL with normal resistance $R_{\rm N}$, heat loss coefficient $k$ and critical current $I_{\rm c}(T_{\rm WL},\Phi)=I_{\rm c}^{\rm 0}(T_{\rm WL})\mid\cos (\pi \Phi/\Phi_{\rm 0})\mid$. The flux-modulation of the critical current alters the expression for the dynamic retrapping current into
\begin{equation}
\frac{I_{\rm r}^{\rm dyn}(\Phi)}{I_{\rm c}^{\rm 0}}=\frac{\sqrt{\sqrt{1+4\beta^2\cos^4(\pi \Phi/\Phi_{\rm 0}) }-1}}{\sqrt{2}\beta \mid\cos (\pi \Phi/\Phi_{\rm 0})\mid}.
\end{equation}
At zero flux, this expression matches with \mbox{Eq. \ref{dynamic}}. In the limit of a small $\beta$, one recovers the usual $\mid\cos (\pi \Phi/\Phi_{\rm 0})\mid$ modulation. In contrast, $I_{\rm h}$ is independent of the flux.

\begin{figure}[t!]
	\includegraphics[width=\columnwidth]{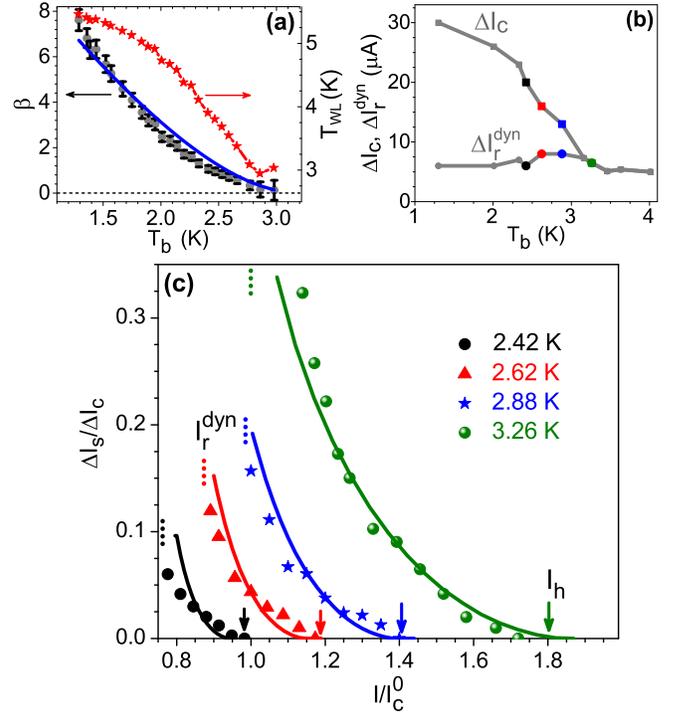}
	\caption{(a) Black symbols: change in $\beta$ with bath temperature for the measured first device as found using $I_{\rm r}^{\rm dyn}$ and $I_{\rm c}^{\rm 0}$ in the hysteretic regime (below $T_h \simeq$ 3 K). The blue solid line is a fit as per Eq. \ref{eq:beta} in DTM with $k$ = 2.6 nW/K. Red symbols and line represent the WL temperature $T_{\rm WL}$ as a function of the bath temperature $T_{\rm b}$ and at a bias current equal to the corresponding $I_{\rm r}^{\rm dyn}$. (b) Variation of the modulation amplitudes $\Delta I_{\rm c}$ and $\Delta I_{\rm r}^{\rm dyn}$ with bath temperature. (c) Symbols: measured supercurrent modulation amplitude in units of the critical current modulation amplitude $\Delta I_{\rm s}/\Delta I_{\rm c}$ in the dissipative state as a function of bias current ranging from $I_{\rm r}^{\rm dyn}$ to $I_{\rm h}$ at different bath temperatures. Solid lines: best fit to the DTM with fit parameters listed in Table I. The value of $I_{\rm c}^{\rm 0}$, setting the x-axis scale, depends on the bath temperature. Arrows and dotted lines indicate the positions of the static and dynamic retrapping current $I_{\rm r}^{\rm stat/dyn}$ respectively.}
	\label{fig:Is-fit}
\end{figure}

For every bath temperature, we use Eq. \ref{dynamic} of the DTM with the measured zero-flux dynamic retrapping current $I_{\rm r}^{\rm dyn}$ and critical current $I_{\rm c}^0$ to extract the value of $\beta$. Figure \ref{fig:Is-fit}(a) shows how the parameter $\beta$ varies with the bath temperature $T_{\rm b}$ over the hysteretic regime, from close to zero at $T_{\rm h}$ to about 8 at 1.3 K. For small $\beta$ values, $I_{\rm r}^{\rm dyn}$ and $I_{\rm c}^{\rm 0}$ are (almost) indistinguishable in IVCs. Thus the error bars in $\beta$ increases when $T_{\rm b}$ is increased towards $T_{\rm h}$, see \mbox{Fig. \ref{fig:Is-fit}(a)} and the method cannot be used beyond. As discussed below, the variation of the residual supercurrent with the bias current is then a more appropriate method to extract the value of $\beta$. The temperature coefficient of the measured critical currrent $I_{\rm c}^{\rm 0}(T_{\rm b})$ below $T_{\rm h}$ being known (see \mbox{Fig. \ref{fig:ivc}(b)}), we use \mbox{Eq. \ref{eq:beta}} with $k$ as the single free parameter to fit the $\beta(T_{\rm b})$ curve. We obtain $k =$ 2.6 nW/K. Alternatively, one can also use the value of $I_{\rm c0}^{\rm 2}R_{\rm N}/kT_{\rm c}=13.9$ obtained from the fit of the dynamic retrapping current as a function of the bath temperature $I_{\rm r}^{\rm dyn}(T_{\rm b})$, which gives k $\approx$ 2.8 nW/K. Using a typical value of heat-transfer coefficient as 5 W/cm$^2$.K,\cite{hazraprb,nikhilprl} and these two close values of $k$, the effective heat loss area is estimated to be $\sim0.06$ $\mu$m$^2$ which is larger than but still comparable to the $0.16\times0.04$ $\mu$m$^2$ area of the WL.

Using the variation of the parameter $\beta$ as a function of the bath temperature $T_{\rm b}$, \mbox{Eq. \ref{temperature_p}} provides us with the behavior of the WL temperature $T_{\rm WL}$ at a bias current equal to the corresponding dynamic retrapping current $I_{\rm r}^{\rm dyn}$, see \mbox{Fig. \ref{fig:Is-fit}(a)}. At the crossover temperature $T_{\rm h}$ where $\beta$ is small, the WL is at thermal equilibrium with bath, \mbox{i.e.} $T_{\rm WL}\approx T_{\rm b}$ as in the isothermal RSJ model. Towards low temperature, the WL temperature $T_{\rm WL}$ increases towards $T_{\rm c}$.

A small $\beta$ value, required for observation of the dynamic regime, necessitates a small critical current $I_{\rm c}$ and/or a large thermal conductance to the bath $k$. Compared to earlier similar devices \cite{nikhilprl} for which we estimate $\beta$ to be about 20 at 4.2 K, we enhanced $k$ by widening the leads right outside the SQUID loop, while still keeping a short and narrow neck between the SQUID loop and the wide leads to avoid vortices between the two WL's current path.\cite{bezryadin} As described earlier, we reduced $I_{\rm c}$ approximately by an order of magnitude by trimming the Nb thickness down, which overall dominates the decrease in $\beta$ value. The value $\beta$ = 2 obtained at 2.1 K, see \mbox{Fig. \ref{fig:Is-fit}(a)}, approximately defines the low temperature limit for practical operation of the SQUID in the voltage-modulation mode, significantly below the hysteresis temperature $T_{\rm h}$. At lower temperatures, the bias-current range of dynamic regime is narrow and the voltage oscillations of small amplitude.

In the experiment, and as discussed above, the critical current is not fully modulated by the flux, which implies the same for the supercurrent. Thus one cannot compare directly the supercurrent calculated from the model to the one deduced from the measured voltage oscillations. We consider the amplitude of the supercurrent modulation in units of the critical current modulation by the flux, \mbox{i.e.} $\Delta I_{\rm s}/\Delta I_{\rm c}$. From the experimental data, we calculate $\Delta I_{\rm s}/\Delta I_{\rm c}$ as being equal to $ \Delta V/(R_{\rm N} \Delta I_{\rm c})$ where $\Delta V$ and $\Delta I_{\rm c}$ are the modulation amplitude of, respectively, the voltage and the critical current. As for the theory, we calculate:
\begin{equation}
\frac{\Delta I_{\rm s}}{\Delta I_{\rm c}}=\frac{I-\sqrt{I^2-{I_{\rm c}^0}^2(T_{\rm WL})}}{I_{\rm c}^0(T_{\rm b})}
\end{equation}
with the temperature $T_{\rm WL}$ being found using \mbox{Eq. \ref{temperature_p}} with $\beta$ as the single adjustable parameter. 

Figure \ref{fig:Is-fit}(c) shows the experimental (symbols) and theoretical (lines) values of the ratio $\Delta I_{\rm s}/\Delta I_{\rm c}$ as a function of the normalized bias current capturing most of the dynamic regime. A very good quantitative agreement is obtained. The fit values of $\beta$ listed in Table \ref{table1} agree well with those deduced, and plotted in \mbox{Fig. \ref{fig:Is-fit}(a)}, from the analysis of the dynamic retrapping current. The comparison made here is fully justified only in the case of a sinusoidal $I_{\rm s}(\varphi)$. Extending this to the case of a non-sinusoidal current-phase relation is intuitive but still not fully theoretically established. Still, the successful comparison of experimental data and theoretical calculation demonstrates that the DTM describes accurately the transition between the isothermal Josephson junction behavior and the electronically-overheated and hysteretic $\mu$-SQUID behavior.

\begin{table}[t!]
	\begin{tabular}{|M{1.5cm}|M{1.5cm}|M{1.5cm}|M{1.5cm}|M{1.5cm}|}
		\hline
		T$_{\rm b}$ (K) & I$_{\rm c}^{\rm 0}$ ($\mu$A) & I$_{\rm r}^{\rm dyn}$ ($\mu$A) & $\beta$ from $I_{\rm r}^{\rm dyn}$ value& $\beta$ from \mbox{Fig. \ref{fig:Is-fit}} fit \\ \hline
		2.42 & 58 & 44 & 1.13 & 1.1 \\
		2.62 & 46 & 40 & 0.66 & 0.74\\
		2.88 & 40 & 38 & 0.35 & 0.5 \\
		3.26 & 31.6 & 31.6 & - & 0.28  \\\hline
	\end{tabular}
	\caption{Sample parameters including the calculated and fitted values of $\beta$ for different temperatures.}
	\label{table1}
\end{table}

\begin{figure}[h!]
	\includegraphics[width=0.7\columnwidth]{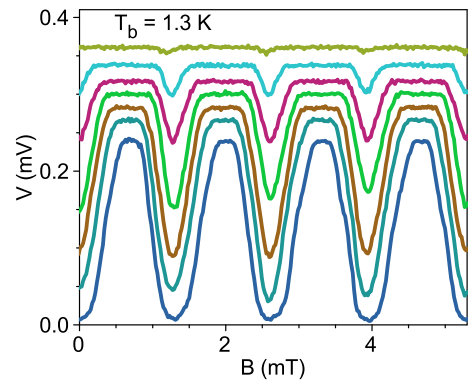}
	\caption{$V-B$ oscillations of another device with a small critical current $I_{\rm c}^{\rm 0}$ = 51 $\mu$A and thus a small $\beta$ = 0.36 at 1.3 K. Here the bias current ranges from 50 to 70 $\mu$A.}
	\label{fig:VBO-1.3K}
\end{figure}

The insights gained from the above study, in particular the key role of the parameter $\beta$, provide a guideline for designing devices with improved performance. While for $0.4<\beta<2$ a wide dynamic regime is obtained, featuring both hysteretic behavior and SQUID voltage-oscillations, one needs to reach $\beta<0.4$ so that hysteresis disappears and the voltage modulations reach a significant fraction of $\Delta/e$. This is illustrated for a device in \mbox{Fig. \ref{fig:VBO-1.3K}}, with narrower WLs as compared to the previous one, resulting in a smaller critical current $I_{\rm c}$ and thus a small $\beta\sim 0.36$, even at 1.3 K.

In this device, the flux-to-voltage transduction function $V_\Phi$ is 1 mV/$\Phi_{\rm 0}$ at 1.3 K. With an estimated voltage noise of 1 nV$/\sqrt{\rm Hz}$ in our circuit, we find a flux noise density $\sqrt{S_\Phi} \approx$ 1 $\mu \Phi_{\rm 0}/\sqrt{\rm Hz}$, \mbox{i.e.} significantly below the values of 3 $\mu \Phi_{\rm 0}/\sqrt{\rm Hz}$ previously reported in non-hysteretic $\mu$-SQUIDs using room temperature amplifiers.\cite{granatasquid,troeman-2007} The use of a low-temperature current amplifier,\cite{mag2} while voltage biasing the $\mu$-SQUID, is expected to further improve the sensitivity.

\section{Conclusion}
In conclusion, we discussed the crossover from the fully overheated WL, \mbox{i.e.} the SBT regime, where the supercurrent is either on or off, to the (isothermal) RSJ case, where the supercurrent contribution decays progressively when the bias current exceeds the critical current. This physics is relevant not only for WLs but also for Josephson junctions based on nanowires, two-dimensional materials and topological insulators, where a large supercurrent density can appear, implying a large power density at the resistive switch, together with poor heat evacuation, thus creating hysteresis. A single parameter $\beta$ reflects the balance between the heat evacuation from the WL and the injected heat, it can be tuned by trimming the critical current and/or varying the thermal coupling to the bath. This balance determines for the amplitude of voltage modulation in the phase dynamic regime. The existence of voltage oscillations makes hysteretic $\mu$-SQUIDs useful as flux-to-voltage transducers for probing magnetism at the nanoscale with a wide bandwidth.

\section*{ACKNOWLEDGEMENTS}
We are indebted to Thierry Crozes for help in the device fabrication at the Nanofab platform at N\'eel Institute. Help from Avijit Duley in calculating the current-phase relation is thankfully acknowledged. SB acknowledges financial support from CSIR, Government of India. AKG acknowledges a research grant from the SERB-DST of the Government of India. AKG also thanks Universit\'e Grenoble Alpes for a visiting position. CW and HC acknowledge financial support from the LabEx LANEF (ANR-10-LABX-51-01) project ÒUHV-NEQÓ. We acknowledge financial support from CEFIPRA through project 5804-2.

$^\dagger$ Institute of Engineering Univ. Grenoble Alpes

\section*{APPENDIX A: ADDITIONNAL EXPERIMENTAL DATA}

\begin{figure}[h!]
	\includegraphics[width=0.85\columnwidth]{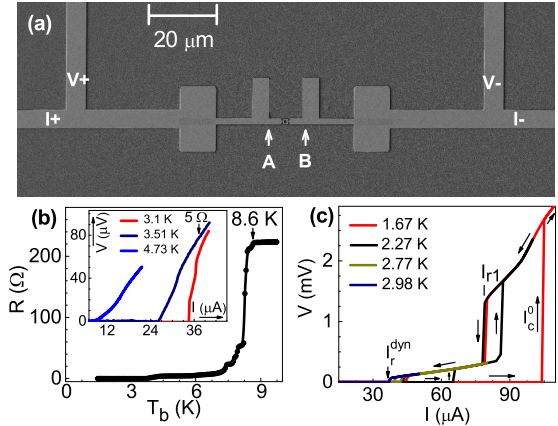}
	\caption{(a) Larger area SEM image of the $\mu$-SQUID showing the current and voltage leads. (b) $R-T_{\rm b}$ plot at a fixed $I$ = 10 $\mu$A showing the superconducting transition at an onset temperature of 8.6 K. Inset shows some IVCs in the non-hysteretic regime. (c) Hysteretic IVCs over larger range of bias current showing thermal instability at $I_{\rm r1}$.}
	\label{fig:sem-res}
\end{figure}

Figure \ref{fig:sem-res}(a,b) shows the large scale SEM image of the $\mu$-SQUID together with its resistance versus bath temperature ($R-T_{\rm b}$) curve and non-hysteretic IVCs. The WL $T_{\rm c}$ of 6 K cannot be found from R-T plot as the bias current of 10 $\mu$A is too high.

Figure \ref{fig:sem-res}(c) shows hysteretic I-V characteristics over a large bias range. We see multiple re-trapping currents with the larger magnitude ones representing thermal instabilities in wider portions (see arrows labeled as A and B in \mbox{Fig. \ref{fig:sem-res}(a)} of the device as evidenced from the resistance values above the respective re-trapping currents. The physics of thermal instability at higher bias currents (above $I_{\rm r1}$), see \mbox{Fig. \ref{fig:sem-res}(c)}, beyond the dynamic regime has been already reported by some of us.\cite{nikhilprl,nikhilsust}
%\textcolor{red}{The relevant parameters for this studied device are summarized in Table \ref{table2}.}

\begin{figure}[h!]
	\includegraphics[width=0.55\columnwidth]{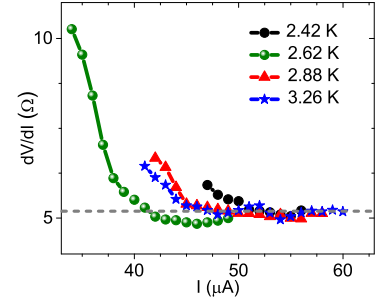}
	\caption{The variation of the differential resistance $dV/dI$ with $I$ in the range $I_{\rm r}^{\rm dyn}<I$. The dotted line shows the saturation to $R_{\rm N}\approx5.2$ K for $I\geq I_{\rm h}$.}
	\label{fig:dvdi}
\end{figure}

\begin{figure}[h!]
	\includegraphics[width=0.9\columnwidth]{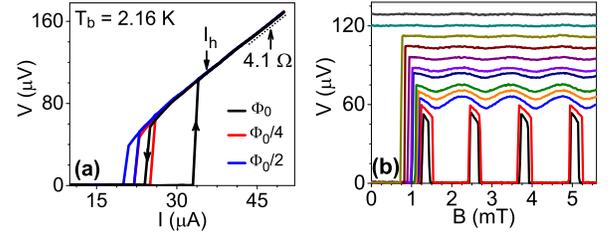}
	\caption{(a) $I_{\rm r}^{\rm dyn}$ oscillation with B at $T_{\rm b}$ = 2.16 K for a third $\mu$-SQUID with $T_{\rm h}$ = 2.75 K. (b) V-B oscillations at this temperature for $I$ = 25 to 38 $\mu$A.}
	\label{fig:anotherset}
\end{figure}

Similar results on transport measurements observed in another (third) $\mu$-SQUID are shown in \mbox{Fig. \ref{fig:anotherset}(a,b)}. The crossover temperature $T_{\rm h}$ of this device is found to be 2.75 K. In the hysteretic regime (below 2.75 K), the retrapping current $I_{\rm r}^{\rm dyn}$ oscillates along with $I_{\rm c}$ with $B$ (see \mbox{Fig. \ref{fig:anotherset}(a)} for $T_{\rm b}$ = 2.16 K). Estimated resistance from the linear ohmic branch of IVCs beyond $I_{\rm h}$ is 4.1 $\Omega$. Period of oscillation is same, \mbox{i.e.} 1.25 mT, as that of the first device. The $V-B$ oscillations at 2.16 K are shown in \mbox{Fig. \ref{fig:anotherset}(b)} for bias currents ranging from $I$ = 25 to 38 $\mu$A.

%\begin{table*}[t!]
%	 \begin{tabular}{M{2cm}M{2cm}M{2cm}M{2cm}M{2cm}M{2cm}M{2cm}M{2cm}}
%		\hline \hline
%		\vspace{0.1cm}
%		WL length (nm) & WL width (nm) & Nb thickness (nm) & Onset $T_{\rm c}$ $(K)$ & $I_{\rm c0}$ ($\mu$A) & $J_{\rm c0}$ ($MA/cm^2$)& $\rho_{\rm 0}$ ($\mu \Omega$-cm) & $k$ ($nW/K$) \\ \hline
%		\vspace{0.1cm}
%		160 & 40 & 20 ($\pm$2) & 8.6 & 211.2 & 33 & 11.6 & $\sim$3  \\\hline
%		\hline
%	\end{tabular}
%	\caption{Summary of the device material parameters for the studied device.}
%	\label{table2}
%\end{table*}

\section*{APPENDIX B: NON-SINUSOIDAL $I_{\rm s}(\varphi)$ RELATION AND $I_{\rm c}$ MODULATION}
\begin{figure}[h!]
	\centering
	\includegraphics[width=\columnwidth]{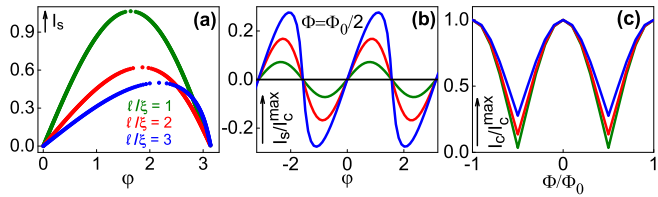}
	\caption{(a) $I_{\rm s}(\varphi)$ relation for different $\ell/\xi$ values. For $\ell/\xi\geq4$, $I_{\rm s}$ is no longer single valued. (b) $I_{\rm s}/I_{\rm c}^{\rm max}$ as a function of $\varphi$ at $\Phi=\Phi_{\rm 0}/2$. Here $I_{\rm c}^{\rm max}$ is SQUID critical current at zero flux. (c) $I_{\rm c}/I_{\rm c}^{\rm max}$ variation with flux. The green, red and blue curves, respectively, represent $\ell/\xi=$1, 2 and 3. The black line in (b) shows $I_{\rm s}$ for perfectly sinusoidal $I_{\rm s}(\varphi)$ relation at $\Phi=\Phi_{\rm 0}/2$.}
	\label{fig:fig1}
\end{figure}

As opposed to the short ($\ell<<\xi$) WLs, where the supercurrent-phase $I_{\rm s}(\varphi)$ relation is sinusoidal ($I_{\rm s}=I_{\rm c}\sin\varphi$),\cite{likharev} the longer WLs ($\ell\geq\xi$) exhibit a non-sinusoidal $I_{\rm s}(\varphi)$. Using Ginzburg-Landau theory (which is valid close to $T_{\rm c}$), the $I_{\rm s}(\varphi)$ relation for a single WL of different lengths is calculated and shown in \mbox{Fig. \ref{fig:fig1}}.\cite{wolf} For a SQUID with identical WLs, the magnetic flux $\Phi$ gives rise to a phase difference $2\pi\Phi/\Phi_{\rm 0}$ between the two WLs. Adding the two WL's supercurrents with this phase difference, we obtain the total supercurrent $I_{\rm s}$. Figure \ref{fig:fig1}(b) shows the total $I_{\rm s}$ as a function of $\varphi$ at $\Phi=\Phi_{\rm 0}/2$. For a short WL-based SQUID, with perfectly sinusoidal $I_{\rm s}(\varphi)$, this total $I_{\rm s}$ at $\Phi=\Phi_{\rm 0}/2$ is identically zero, which is clearly not the case for long WL based SQUIDs.

Figure \ref{fig:fig1}(c) shows the $I_{\rm c}(\Phi)$ oscillations for $\ell/\xi=$1, 2 and 3 with $I_{\rm c}^{\rm max}$ as the maximum (with respect to $\varphi$) value of $I_{\rm s}$ at a given $\Phi$. The $I_{\rm c}$ modulation amplitude decreases with increasing $\ell/\xi$. This demonstrates how the $I_{\rm c}$ modulation of a SQUID with flux is limited by the non-sinusoidal $I_{\rm s}(\varphi)$ relation of the WLs.

\section*{APPENDIX C: RCSJ MODEL}
Here we attempt the fitting of the super-current relative modulation $\Delta I_{\rm s}/\Delta I_{\rm c}$ with the RCSJ model. This is anyway not very plausible due to the lack of sharp cut-off in $I_{\rm s}$ which is quite apparent in the experiments.

According to the RCSJ model,\cite{squidbook} the current $I$, as shared between resistances, capacitances and the Josephson junctions of the SQUID, can be written as:
\begin{equation}
I=I_{\rm c}^{\rm 0}\sin\varphi+\frac{\mathcal{V}(t)}{R_{\rm N}}+C\frac{d\mathcal{V}(t)}{dt}.
\end{equation}
Here, $2R_{\rm N}$ and $C/2$ are the resistance and capacitance of each of the two junctions. We have again assumed the screening parameter $LI_{\rm c}^{\rm 0}/\Phi_{\rm 0}<<1$. using the same dimensionless quantities ate Section I, we get
\begin{equation}
i=\sin\varphi+\dot{\varphi}+\beta_c\ddot{\varphi}.
\label{eq:rcsj}
\end{equation}
Here $\beta_{\rm c}=\frac{2\pi}{\Phi_{\rm 0}}I_{\rm c}^{\rm 0}R^2_{\rm N}C$ is the effective Stewart-McCumber parameter for the SQUID.

\begin{figure}[h!]
	\includegraphics[width=0.7\columnwidth]{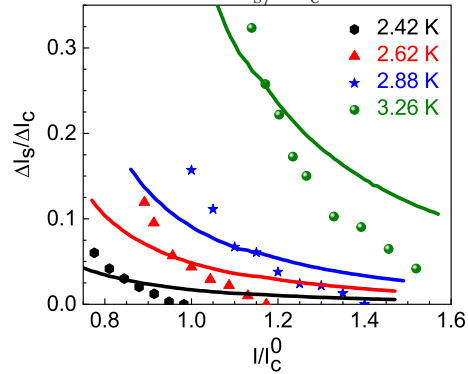}
	\caption{Variation of residual super-current ($\Delta I_{\rm s}/\Delta I_{\rm c}$) in the resistive branch at $T_{\rm b}$ = 2.42, 2.62, 2.88 and 3.26 K as calculated from experimental data (shown by symbols). Dotted lines represent the best fit to the RCSJ model. $I_{\rm c}^0$ in x-axis is the critical current value at zero flux for respective $T_{\rm b}$.}
	\label{fig:rcsj-fit}
\end{figure}

\begin{table}[b!]
	\begin{tabular}{|M{1.5cm}|M{1.5cm}|M{1.5cm}|M{1.5cm}|M{1.5cm}|}
		\hline
		T$_{\rm b}$ (K) & I$_{\rm c}^{\rm 0}$ ($\mu$A) & I$_{\rm r}^{\rm dyn}$ ($\mu$A) & calculated $\beta_{\rm c}$ & fitted $\beta_{\rm c}$  \\ \hline
		2.42 & 58 & 44 & 2.13 & 5.4 \\
		2.62 & 46 & 40 & 1.5 & 3.2\\
		2.88 & 40 & 38 & 0.8 & 2.33 \\
		3.26 & 31.6 & 31.6 & - & 1.0  \\\hline
	\end{tabular}
	\caption{Comparison of measured $I_{\rm r}^{\rm dyn}$, $I_{\rm c}^{\rm 0}$ with calculated (from $I_{\rm r}^{\rm dyn}$ data) and fitted $\beta_{\rm c}$ for different temperatures.}
	\label{table3}
\end{table}

We get the IVCs from the numerical steady-state solutions of Eq. \ref{eq:rcsj} and $V=\Phi_{\rm 0}\dot{\varphi}/2\pi$. Using these solutions, we find that $I_{\rm r}^{\rm dyn}/I_{\rm c}^{\rm 0}$ depends on $\beta_{\rm c}$. This fact is used to extract the $\beta_{\rm c}^{\rm calc}$ values for experimentally measured $I_{\rm r}^{\rm dyn}/I_{\rm c}^{\rm 0}$. Using the expression $I_{\rm s}=I-V/R_{\rm N}$, we have tried to fit the measured $\Delta I_{\rm s}/\Delta I_{\rm c}$ to the RCSJ model in Fig. \ref{fig:rcsj-fit}. The fitted $\beta_{\rm c}$ values are listed in Table \ref{table3} together with the values extracted from the values of the dynamic retrapping current $I_{\rm r}^{\rm dyn}$. We see from the fit that RCSJ does not fit well as compared to the DTM.


\begin{thebibliography}{30}
\bibitem{notation} Note that in the original paper introducing the DTM,\cite{anjanjap} we used a different notation, $I_{\rm r}^{\rm stat}$, for this current.
\end{thebibliography}
\end{document}